\documentclass[%
 aip,
 amsmath, 
 amssymb,
floatfix,
 reprint %
]{revtex4-1}
\usepackage{graphicx}
\usepackage{dcolumn}
\usepackage{bm}
\usepackage[utf8]{inputenc}
\usepackage[T1]{fontenc}
\usepackage{mathptmx}
\usepackage{etoolbox}
\usepackage{booktabs}
\usepackage{widetable}
\usepackage{threeparttable}
\usepackage[version=3]{mhchem}
\makeatletter
\def\@email#1#2{%
 \endgroup
 \patchcmd{\titleblock@produce}
  {\frontmatter@RRAPformat}
  {\frontmatter@RRAPformat{\produce@RRAP{*#1\href{mailto:#2}{#2}}}\frontmatter@RRAPformat}
  {}{}
}%
\makeatother

\usepackage{hyperref}
\usepackage{placeins}
\usepackage{xcolor}

\definecolor{aaltoOrange}{RGB}{255,121,0}%

\newcommand{\DG}[1]{\textcolor{black}{{#1}}}

\DeclareMathOperator\Tr{Tr}
\DeclareMathOperator\ii{i}

\def\argmin{\qopname\relax m{argmin}}
\begin{document}


\title{Benchmarking the accuracy of the separable resolution of the identity approach for correlated methods in the numeric atom-centered orbitals framework}

\author{Francisco A. Delesma}
\email{francisco.delesma@aalto.fi}
\affiliation{Department of Applied Physics, Aalto University, FI-02150 Espoo, Finland}

\author{Moritz Leucke}
\affiliation{Faculty for Chemistry and Food Chemistry, Technische Universit\"at Dresden, 01062 Dresden, Germany}

\author{Dorothea Golze}
\affiliation{Faculty for Chemistry and Food Chemistry, Technische Universit\"at Dresden, 01062 Dresden, Germany}

\author{Patrick Rinke}
\affiliation{Department of Applied Physics, Aalto University, FI-02150 Espoo, Finland}

\date{\today}

\begin{abstract}
Four-center two-electron Coulomb integrals routinely appear in electronic structure algorithms. The resolution-of-the-identity (RI) is a  popular technique to reduce the computational cost for the numerical evaluation of these integrals in localized basis-sets codes.
Recently, Duchemin and Blase proposed a separable RI  scheme [J. Chem. Phys. 150, 174120 (2019)], which preserves the accuracy of the standard global RI method with the Coulomb metric (RI-V) and permits the formulation of cubic-scaling random phase approximation (RPA) and $GW$ approaches. Here, we present the implementation of a separable RI scheme within an all-electron numeric atom-centered orbital framework. We present comprehensive benchmark results using the Thiel 
\DG{and the GW100 test set.} Our benchmarks include atomization energies from Hartree-Fock, second-order M{\o}ller–Plesset (MP2), coupled-cluster singles and doubles, RPA and renormalized second-order perturbation theory as well as quasiparticle energies from $GW$. We found that the separable RI approach reproduces RI-free HF calculations within 9 meV and MP2 calculations within 1 meV. \DG{We have confirmed that the separable RI error is independent of the system size by including disordered carbon clusters up to 116 atoms in our benchmarks.}
\end{abstract}

\maketitle
 
\section{Introduction}
\label{sec:introduction}
The accurate computation of molecular or materials' properties that derive from the electronic structure is important in quantum chemistry and materials science. Kohn-Sham density functional theory \cite{HK64,KS65} (KS-DFT) is the most popular electronic-structure method to date, providing a reasonable compromise between accuracy and computational efficiency.\cite{Jones2015} However, KS-DFT faces several limitations, including the correct description of strong correlation,\cite{Burke2012} van-der-Waals (vdW) interactions\cite{Burke2012,RPA_Ren12} or excited-state properties.\cite{GWcompe} Various methods have been explored to overcome or mitigate the short-comings of KS-DFT. 
They can be broadly classified into wave-function based approaches, such as coupled cluster (CC)\cite{cc_COESTER1960477,cc_cizek} and many-body perturbation theory (MBPT) schemes, \DG{although the distinction is becoming blurry as MBPT reformulations of CC methods were presented in the literature.\cite{Nooijen1992,RPA_Ren12,Scuseria2013,Lange2018}}
Examples for MBPT schemes are second order M{\o}ller-Plesset (MP2) perturbation theory,\cite{mp234} the random phase approximation\cite{Bohm1953,GellMann1957} (RPA),
the $GW$ approximation to Hedin's  equations,\cite{Hedin65} renormalized second-order perturbation theory (rPT2), which  includes next to RPA the renormalized single excitation \cite{Ren_11_SE} (rSE) and second-order screened exchange\cite{sosex09,sosex10} (SOSEX).

Four-center two-electron repulsion integrals (4c-ERIs) appear in all of the mentioned methods and their direct evaluation scales ${O}(N^4)$ with respect to basis set size $N$ and thus with system size. Resolution-of-the-identity (RI), also known as density fitting, is a popular technique to alleviate the computational demands for the calculation of the 4c-ERIs within a localized-basis-set framework,\cite{Whitten73,Dunlap79,Weigend98,Weigend02} which is typically used in the quantum chemistry community. RI expands the product of two atomic orbitals (AOs), $\varphi_i\varphi_j$, in a linear combination of auxiliary basis functions (ABFs) and replaces the exact evaluation of the four-center Coulomb interactions with the Coulomb interactions of a set of ABFs. The most popular and accurate RI flavor is the global Coulomb-fitting scheme RI-V, where the 4c-ERIs are refactored into two- and three-center Coulomb integrals.\cite{Vahtras93} Even though the number of ABFs is at least two times larger than the corresponding AO basis sets,\cite{Weigend98,Weigend02} the refactoring reduces the memory demand and the number of required operations considerably. The RI integrals can be evaluated analytically or numerically depending on the AO type. Analytic schemes are used for, e.g., Gaussian-type orbitals (GTOs),\cite{Obara1986,Ahlrichs2006,Asadchev2023,Giese2008,Golze2017a,Kovacevic2022} while integrals over numeric atom-centered orbitals (NAOs) are computed on grids.\cite{Ren12}

Alternative approaches to approximate the 4c-ERIs employ partial or full real-space quadrature strategies, regardless of the AO type. Examples are Friesner's pseudospectral method,\cite{Pseudospectral} which involves the use of both numerical grids and analytical integrals, and the related semi-numerical integration scheme developed by Neese and coworkers.\cite{NEESE200998} Another alternative approach to RI is the least square tensor hypercontraction (LS-THC) method,\cite{LTHC,Hohenstein2012} which completely expresses the 4c-ERIs on real-space grids, introducing a high complexity of $O(N^{4-5})$.
In 2015, a cubic-scaling reformulation of the THC method was proposed,\cite{Lu2015} which was later referred to as interpolative separable density fitting (ISDF) in the literature.\cite{Lu2017,Hu2017,Qin2023} As in RI schemes, ISDF linearly expands the AO pairs $\varphi_i\varphi_j$, but uses a separable expression for the expansion coefficients. While the RI expansion coefficients entangle $\varphi_i(\mathbf{r})\varphi_j(\mathbf{r})$ through the three-center integrals, the ISDF coefficients are defined as the product $\varphi_i(\mathbf{r}_k)\varphi_j(\mathbf{r}_k)$, where $\{\mathbf{r}_k\}$ is a subset of the real-space grid $\{\mathbf{r}\}$. The so-called interpolation points $\{\mathbf{r}_k\}$ are selected with randomized QR decomposition or K-means clustering algorithms and the ISDF-equivalent to the ABFs are then obtained by least-square optimization.\cite{Qin2023} 

The ISDF method has been used to accelerate different electronic structure methods, including HF or hybrid functionals,\cite{Hu2017,Dong2018,Qin2020a,Qin2020b,Wu2022} MP2,\cite{Lee2020} RPA,\cite{Lu2017} $GW$,\cite{Gao2020} the Bethe-Salpeter equation\cite{Hu2018,Henneke_ISDF_2020} (BSE) or Quantum Monte Carlo,\cite{Malone2019} and has been implemented in plane-wave, real-space, GTO and NAO basis set codes.\cite{Qin2023} 
Recently, Duchemin and Blase\cite{BlaseRPA} proposed a separable real-space RI scheme (RI-RS), where ISDF-like expansion coefficients were obtained over a compact set of real-space points by fitting them to the RI-V coefficients.  The RI-RS approach does not reduce the computational cost of canonical implementations. However, Duchemin and Blase showed that it enables the formulation of highly accurate cubic-scaling RPA \cite{BlaseRPA} and $GW$\cite{BlaseGW} algorithms, in combination with the Laplace transform technique\cite{Almloef1991} and the space-time method,\cite{Rojas1995} respectively.

In the last decade, RPA and $GW$ became popular in quantum chemistry\cite{GWcompe} and low-scaling algorithms for RPA\cite{Wilhelm16,BlaseRPA,RPA_Graf_low,Drontschenko2022,Shi2023} and $GW$\cite{Wilhelm18,tsgw21,BlaseGW,Foerster2020,Foerster2021a,Foerster2023,PanadesBarrueta2023,Graml2023} are currently very actively developed for localized basis sets. Sparsity has been introduced in the low-scaling RPA and $GW$ expressions in different ways. One approach is to replace the Coulomb metric in  RI-V with a local metric, such as an overlap,\cite{Wilhelm16,Wilhelm18} attenuated\cite{RPA_Graf_low} and truncated\cite{tsgw21,Graml2023} Coulomb metric. Alternatively, the Coulomb metric is kept, but  adopted to a local RI scheme,\cite{Shi2023,Foerster2020,Foerster2021a,Foerster2021b,Foerster2023} which restricts the ABFs to the centers of the AO pairs. High accuracy compared to canonical implementations has been achieved with the first approach,\cite{tsgw21} at the price of introducing sparse tensor-tensor multiplications, which require special libraries. The second approach, local RI, is computationally less demanding than the first scheme, but controlling the accuracy can be more challenging.\cite{Spadetto2023} The RI-RS approach now offers a third way. Its advantage, compared to the other two approaches, is its automatic sparsity in the RPA polarizability. In addition, sparse tensor operations are avoided and RI-RS is highly accurate as previous work suggests.\cite{BlaseRPA} 

Duchemin and Blase implemented the RI-RS approach within a GTO framework and applied it to HF, MP2, RPA and $GW$.\cite{BlaseRPA,BlaseGW} In this work, we present our implementation and validation of RI-RS within an NAO framework and extend the use of the RI-RS approach to coupled cluster singles and doubles (CCSD), SOSEX, rSE  and rPT2 energies. The purpose of this paper is to validate the accuracy of the RI-RS scheme for a wide range of methods and highlight its potential for deriving low-scaling algorithms.

The article is organized as follows. In Section~\ref{sec:theory}, we briefly review the theoretical background of the electronic-structure methods and the numerical techniques used. In Section~\ref{sec:implementation} we present details of the implementation. Computational details are given in Section~\ref{sec:computational_details}. In Section~\ref{sec:results}, we discuss comprehensive benchmark results, including total, atomization and quasiparticle energies. Finally, we draw conclusions in Section~\ref{sec:conclusion}.

\section{Theory}
\label{sec:theory}
In this section, we summarize the RI-V formalism and the RI-V working equations for the HF exchange and the CCSD, MP2, RPA correlation energies. We also briefly review the theoretical background of the rPT2 and $G_0W_0$ methodologies and their corresponding RI-V expressions. We then introduce the RI-RS scheme in detail. We restrict the presentation in Section~\ref{sec:theory} to the closed-shell case and drop the spin index. Throughout this discussion, we adopt the following indexing: $p, q, ...$ denote general molecular orbitals (MOs), $m, n, ...$  occupied (occ) MOs and $a, b, ...$ unoccupied or equivalently virtual (virt) MOs. We use $i,j,..$ to index the functions of the primary basis set and $P,Q,...$ (in capital letters) are used to denote the ABFs. 

\subsection{Resolution of the identity}

The common ingredient of all the methods discussed in this work are the 4c-ERIs, which are defined as
\begin{equation} \label{4cmos1}
    \left( p q | s t\right) = \iint \frac{\psi_{p}(\mathbf{r})\psi_{q}(\mathbf{r}) \psi_{s}(\mathbf{r}') \psi_{t}(\mathbf{r}')}{|\mathbf{r}-\mathbf{r}'|} d\mathbf{r}d\mathbf{r}',
\end{equation}
where $\psi_{p}(\mathbf{r})$ are MO (also called  single-particle orbitals). The MOs
are expanded in terms of the basis functions $\{\varphi_{i}(\mathbf{r})\}$
\begin{equation}\label{aoexp}
    \psi_{p}(\mathbf{r}) = \sum_{i} c_{p}^{i} \varphi_{i}(\mathbf{r})
\end{equation}
where $c_{p}^{i}$ are the MO coefficients. Inserting the expansion of Eq.\eqref{aoexp} into the Eq.\eqref{4cmos1} yields 
\begin{equation}\label{4cmos2}
    \left( p q | s t \right) = \sum_{ijkl} \left( i j | k l \right) c_{p}^{i} c_{q}^{j} c_{s}^{k} c_{t}^{l}
\end{equation}
where the two-electron integrals in the AO basis 
are defined as 
\begin{equation}\label{4caos1}
    \left( i j | k l \right) = \iint \frac{\varphi_{i}(\mathbf{r})\varphi_{j}(\mathbf{r}) \varphi_{k}(\mathbf{r}') \varphi_{l}(\mathbf{r}')}{|\mathbf{r}-\mathbf{r}'|} d\mathbf{r}d\mathbf{r}'.
\end{equation}

The RI approximation expands the product of two AO functions, $\varphi_{i}(\mathbf{r})\varphi_{j}(\mathbf{r})$, in terms of a set of ABFs, namely

\begin{equation}\label{ridef}
\rho_{ij}(\mathbf{r})\equiv\varphi_{i}(\mathbf{r})\varphi_{j}(\mathbf{r}) \approx \sum_{\mu} C_{ij}^{\mu}P_{\mu}(\mathbf{r}) \equiv \tilde{\rho}_{ij}(\mathbf{r})
\end{equation}
where $\mu$ labels the ABFs, and $\{P_{\mu}\}$
and $C_{ij}^{\mu}$ are the RI expansion coefficients. 

The different RI techniques can be grouped into global and local. In global RI, the sum in Eq.~\eqref{ridef} runs over all ABFs,\cite{Vahtras93} while it is restricted to a subset in local RI. \cite{Baerends1973,Merlot2013,Ihrig2015,Golze2017b} Given the pair $\varphi_i\varphi_j$, where $\varphi_i$ is centered at atom $A$ and $\varphi_j$ at atom $B$, only RI basis functions with centers at $A$ and $B$ are used in local RI expansions. We note that we use global RI in this work and thus exclude local RI from the discussion in the following.

Inserting Eq.~\eqref{ridef} into Eq. \eqref{4caos1} yields the expression
\begin{equation}\label{4cmos3}
    (ij|kl)_{\text{RI}} \equiv \sum_{\mu\nu}C_{ij}^{\mu}V_{\mu\nu}C^{\nu}_{kl}
\end{equation}
with the two-center Coulomb integrals 
\begin{equation}\label{2c-eris}
     V_{\mu\nu}= \iint \frac{{P}_{\mu}(\mathbf{r})P_{\nu}(\mathbf{r}')}{|\mathbf{r}-\mathbf{r}'|}d\mathbf{r} d\mathbf{r}' 
\end{equation}
Several ways to obtain the expansion coefficients, $C_{ij}^{\mu}$, were proposed in the literature.\cite{Vahtras93}  One choice is to use the so-called overlap metric, i.e., to minimize the residual $\delta \rho_{ij}=\tilde{\rho}_{ij}(\mathbf{r})-{\rho}_{ij}(\mathbf{r})$ with respect to $C_{ij}^{\mu}$ by a simple least-square fit, yielding the RI-SVS approximation.\cite{Vahtras93}
Another criterion for obtaining $C_{ij}^{\mu}$ is to use the Coulomb metric, which implies minimizing the Coulomb repulsion of the density residual,
$(\delta \rho_{ij}|\delta \rho_{ij})$, yielding\cite{Whitten73,Dunlap79,Mintmire82,Vahtras93} 

\begin{equation}\label{coeff_riv}
C_{ij}^{\mu}=\sum_{\nu} (ij | \nu )V_{\nu\mu}^{-1}
\end{equation}
where the three-center Coulomb integrals in Eq.~\eqref{coeff_riv} are given by
\begin{equation}\label{3c-eris}
(ij | \nu ) = \iint \frac{\varphi_{i}(\mathbf{r})\varphi_{j}(\mathbf{r})P_{\nu}(\mathbf{r}')}{|\mathbf{r}-\mathbf{r}'|}d\mathbf{r} d \mathbf{r}'.
\end{equation}
Computing the expansion via Eq.~\eqref{coeff_riv} is known as the RI-V approach.  RI-V converges faster with respect to the size of the RI basis set than RI-SVS and is typically more robust and accurate than the latter. \cite{Whitten73,Dunlap79,Mintmire82,Vahtras93} We therefore choose RI-V as reference for the RI-RS coefficients.

Making use of the symmetry of $\mathbf{V}$ defined in Eq.~\eqref{2c-eris}, we can rewrite Eq. \eqref{4cmos3} as
\begin{equation}\label{4cmos4}
    (ij|kl)_{RI} \equiv \sum_{\mu} M_{ij}^{\mu}M^{\mu}_{kl}
\end{equation}
with
\begin{equation}\label{fcao1}
M_{ij}^{\mu}=\sum_{\nu}C_{ij}^{\nu}V_{\nu\mu}^{1/2}.
\end{equation}
Inserting Eq.~\eqref{4cmos4} into Eq.~\eqref{4cmos1}, we obtain the RI reformulation of the 4c-ERIs 
\begin{equation}
     \left( p q | s t \right)_{\text{RI}} = \sum_{\mu} O_{p q}^{\mu}O^{\mu}_{s t}
\end{equation}
using the MO transformation of $M_{ij}^{\mu}$
\begin{equation}\label{fcmo1}
    O_{p q}^{\mu} =\sum_{ij}c_{p}^{i}c_{q}^{j}M_{ij}^{\mu}.
\end{equation} 

\subsection{RI-V expression for different theoretical methods}
\label{sec:riv_methods}
\subsubsection{Hartree-Fock theory}

The HF method is fundamental to electronic structure theory. In HF, the electronic structure is described by a single Slater determinant and the energy is minimized by varying the one-electron orbitals using self-consistent field (SCF) algorithms. Evaluating the exact-exchange energy, $E_x^{\text{HF}}$, is the most expensive step in HF. The RI-V expression of $E_x^{\text{HF}}$ is given by

\begin{equation}\label{Fock_ener}
    E_x^{\text{HF}} = - \sum_{mn}^{\text{occ}}\sum_{\mu}O_{mn}^{\mu}O_{nm}^{\mu}.
\end{equation}

The exact exchange energy is recomputed in each SCF iteration. In practice, we first compute and store the three-center quantity $M_{ij}^{\mu}$ before the start of the SCF loop. In each iteration, the MO coefficients are updated and $M_{ij}^{\mu}$ is transformed to the MO basis using Eq.~\eqref{fcmo1}, which then yields $O_{mn}^{\mu}$. The scaling of Eq.~\eqref{Fock_ener} formally remains  $O(N^4)$,\cite{Ren12} but is in practice usually lower. The dominating step, in particular in our NAO scheme, is the evaluation of the three-center AO integrals (Eq.~\eqref{3c-eris}), which scales $O(N^3)$ for small to medium-sized systems and $O(N^2)$ for large systems because the product $\varphi_i\varphi_j$ becomes sparse.

\subsubsection{M{\o}ller-Plesset perturbation theory}

MP2\cite{mp234} \DG{is widely used in quantum chemistry } 
for adding correlation to the HF ground state.\cite{Cremer2011}
The MP2 correlation energy is obtained by adding a small perturbation to the HF Hamiltonian and then terminating the perturbative expansion at second order. The RI-V expression of the MP2 correlation energy for closed shell systems reads\cite{Feyereisen93,Weigend98,Weigend02}

\begin{eqnarray}\label{mp21}
    E_c^{\textnormal{MP2}}&=&   \sum_{mn}^{\text{occ}}\sum_{ab}^{\text{virt}}\left(\sum_{\mu}O_{ma}^{\mu}O_{nb}^{\mu}\right) \times \nonumber \\ & & \left[\frac{2\left(\displaystyle\sum_{\mu}O_{am}^{\mu}O_{bn}^{\mu}\right) - \left(\displaystyle\sum_{\mu}O_{bm}^{\mu}O_{an}^{\mu}\right)}{\epsilon_{m}+\epsilon_{n}-\epsilon_{a}-\epsilon_{b}} \right]    ,
\end{eqnarray}
where $\varepsilon_{m/n}$ and $\varepsilon_{a/b}$ are the orbital energies of the occupied and  unoccupied states, respectively. The first term in Eq.~\eqref{mp21} is the second-order Coulomb term, which is also referred to as direct MP2 term, and corresponds, in a diagrammatic representation, to the leading term in RPA. The second one is the second-order exchange energy, which appears in a dressed form in SOSEX.\cite{RPA_Ren12}
In addition to the established $O(N^5$) MP2 RI-V scheme, different RI versions have recently been explored to scale MP2 to larger systems.\cite{MP2_Werner03,MP2_Maschio07,MP2_DelBen12,MP2_Glasbrenner20}

\subsubsection{Coupled cluster theory}

Coupled-cluster theory\cite{cc_COESTER1960477,cc_cizek,cc_purvis} is a wave-function-based approach, which is well-established in quantum chemistry\cite{Bartlett2007} and is becoming increasingly popular in computational materials science.\cite{CC_Zhang19,Gruneis2018,Gruneis2023} CCSD with perturbative triples, CCSD(T), is considered as one of the most accurate methods, which is still computationally affordable for larger molecules.\cite{ccsdfhi} However, we will restrict the discussion to CCSD since our goal is the numerical validation of RI-RS rather than providing the best possible computational reference.   

The CCSD wave function is given by an exponential ansatz
\begin{equation}
    |\Psi_{\textnormal{CCSD}}\rangle = e^{\hat{T}}|\Psi_{\textnormal{HF}}\rangle
    \label{cc_ansatz}
\end{equation}
where $\hat{T}$ is the cluster operator, which generates a linear combination of excited determinants when acting on the  HF single Slater determinant ground state. For CCSD, $\hat{T}$ is defined as
\begin{equation}
    \hat{T}=\hat{T_1} + \hat{T}_2
\end{equation}
where $\hat{T}_1$ generates singly- and $\hat{T}_2$ doubly excited determinants. In second quantization these operators read for closed shell systems:
\begin{align}
    \hat{T}_1 &= \sum_{m}\sum_{a}t_{m}^{a}\hat{E}_{m}^{a}\label{cc_t1}\\
    \hat{T}_2 &= \frac{1}{2}\sum_{mn}\sum_{ab}t_{mn}^{ab}\hat{E}_{m}^{a}\label{cc_t2}
\end{align}
The operator $\hat{E}_{m}^{a}$ contains the annihilation operator $\hat{a}$ and the creation operator $\hat{a}^{\dagger}$:
\begin{equation}
    \hat{E}_{p}^{q} = \hat{a}_{p}^{\dagger}(\alpha)\hat{a}_{q}(\alpha) + \hat{a}_{p}^{\dagger}(\beta)\hat{a}_{q}(\beta)
\end{equation}
where $\alpha$ and $\beta$ indicate different spin functions. $t_{m}^{a}$ and $t_{mn}^{ab}$ are the cluster amplitudes, which are obtained by iteratively solving the CCSD equations. The latter are derived by inserting  Eq.~\eqref{cc_ansatz} into the Schrödinger equation and multiplying on the left by $\langle\Psi_{\textnormal{HF}}|e^{-\hat{T}}$,  $\langle\Psi_{m}^{a}|e^{-\hat{T}}$ and  $\langle\Psi_{mn}^{ab}|e^{-\hat{T}}$, where $\Psi_{m}^{a}$ and $\Psi_{mn}^{ab}|$ are singly and doubly excited determinants. 

Once the cluster amplitudes $t_{m}^{a}$ and $t_{mn}^{ab}$ are computed, the closed shell CCSD correlation energy in its RI-V expression is given by~\cite{ccsdfhi} 
\begin{eqnarray}
    E_c^{\textnormal{CCSD}} &=&  \sum_{mn}^{\text{occ}}\sum_{ab}^{\text{virt}} \left[2\left(\sum_{\mu}O_{ma}^{\mu}O_{nb}^{\mu}\right)-\left(\sum_{\mu}O_{mb}^{\mu}O_{na}^{\mu}\right)\right]\times \nonumber \\ && \left(t_{mn}^{ab} + t_{m}^{a}t_{n}^{b}\right)
\end{eqnarray}
The complexity of the described approach is $O(N^6)$, rendering CCSD also a computationally costly approach. 

\subsubsection{Random phase approximation}

The total RPA energy is given as sum of the HF energy and the RPA correlation energy $E_c^{\text{RPA}}$, which corresponds to the fifth rung of ``Jacobs ladder''\cite{Perdew2001} in the context of KS-DFT, where LDA, GGA, meta GGAs and hybrid functionals are placed on rungs 1 to 4 in that order.
In practice, RPA is typically performed on-top of a preceding KS-DFT calculation. This implies that we insert the KS-DFT orbitals and energies in the expressions for the exact HF exchange and $E_c^{\text{RPA}}$. In this post-processing mode, the total RPA energy is obtained by adding both terms to the KS-DFT total energy and subtracting the KS-DFT exchange-correlation contribution.

In a diagrammatic description, RPA expands the direct MP2 term to infinite order. Its RI-V expression reads for the closed-shell case\cite{Eshuis10,Ren12,brunevalmolgw,BlaseRPA}
\begin{equation}
    E_c^{\text{RPA}}=\frac{1}{2\pi}\int_{0}^{\infty} d\omega \Tr{[\ln{(\textbf{1}-\boldsymbol{\Pi}(\ii\omega))} + \boldsymbol{\Pi}(\ii\omega)]}
\end{equation}
with the irreducible polarizability in the ABFs representation\cite{Ren12}
\begin{equation}\label{pi_def}
    \Pi(\text{i}\omega)_{\mu\nu} = 2 \sum_{m}^{\text{occ}} \sum_{a}^{\text{virt}}  \left(\frac{O_{ma}^{\mu}O_{am}^{\nu}}{\text{i}\omega -\epsilon_{a} + \epsilon_{m}} + \text{c.c.} \right)
\end{equation}
where $\varepsilon_{m}$ are again the KS-DFT orbital energies of the occupied states and $\varepsilon_{a}$ are the orbital energies of the unoccupied states, and c.c. denotes the \textit{complex conjugate}. The RI-V reformulation of RPA reduces the scaling from ${O}(N^6)$ to ${O}(N^4)$,\cite{Eshuis10,Eshuis12,Ren12} which still restricts the accessible system sizes to less than 100 atoms. Several scaling reductions have, however, been proposed. \cite{DelBen15,Wilhelm16,RPA_Graf_low,BlaseRPA,Drontschenko2022,Shi2023,Spadetto2023} 

\subsubsection{Renormalized second-order perturbation theory}

While long-range interactions are well described in RPA, short-range exchange and correlation are not sufficiently captured \cite{Singwi1968,RPA_Ren12}  and several schemes have been developed to go beyond RPA. Ren \textit{et al.}\cite{Ren_11_SE} proposed to add the rSE correction to RPA, while Grüneis and coworkers\cite{sosex09} put forward an RPA + SOSEX approach. Adding both corrections (RPA + SOSEX + rSE) was investigated by Paier \textit{et al.},\cite{rPT2_1} and later promoted as renormalized second-order perturbation theory (rPT2).\cite{rPT2Ren13} Benchmark studies\cite{rPT2_1,rPT2Ren13} of small inorganic and organic molecular systems showed that rPT2 is generally superior to RPA+SOSEX or RPA+rSE for reaction barrier heights and atomization energies. It has also been argued that rPT2 is a well-rounded approach from a diagrammatic perspective. The goldstone diagrams of RPA, SOSEX and rSE are distinct infinite summations, where each series sums up topologically similar diagrams. The leading terms, i.e., the first terms in the sum, are the second-order direct, second-order exchange and single excitation terms, respectively. The combination of the three leading terms is known as second-order Rayleigh-Schrödinger perturbation theory (RSPT). This is reflected in the name of the method: the infinite expansion in RPA+SOSEX+rSE can be understood as a renormalization of RSPT yielding the ``renormalized second-order perturbation theory'' or in short rPT2. 

SOSEX originates from the wave-function theory and was first derived in the context of CC.\cite{sosex77,sosex10} Diagrammatically, we add higher order exchange diagrams to RPA in the RPA+SOSEX scheme, which removes the correlation part of the one-electron self-interaction from the theory. The RI-V expression for the SOSEX correlation is given for the closed-shell case as\cite{rPT2Ren13}
\begin{eqnarray}
\label{eq:sosex}
E_c^{\text{SOSEX}} & = & -\frac{1}{2\pi}\int_0^{\infty} d\omega \sum_{mn}^{\text{occ}}\sum_{\text{ab}}^{\text{virt}}\left[ \left(\sum_{\mu}O_{ma}^{\mu}O_{nb}^{\mu}\right) \right. \times  \nonumber  \\ && \left.\left(\sum_{\nu\gamma}O_{ma}^{\nu}\mathcal{E}^{-1}_{\nu\gamma}(\text{i}\omega)O_{nb}^{\gamma}\right)\right] \times \nonumber\\ && 
\mathcal{F}_{ma}(\text{i}\omega)\mathcal{F}_{nb}(\text{i}\omega)
\end{eqnarray}
with the factors

\begin{equation}
    \mathcal{F}_{ma}(\text{i}\omega)=\frac{2(\epsilon_{m} -\epsilon_{a})}{[(\epsilon_{m} -\epsilon_{a})^2 + \omega^2] }
\end{equation}
and the coupling-constant-averaged dielectric function is defined as
\begin{equation}\label{ccadie}
    \mathcal{E}^{-1}(\text{i}\omega)=\int_{0}^{1} d \lambda [\textbf{1}-\lambda{\boldsymbol\Pi}(\text{i}\omega)]^{-1}\lambda 
\end{equation}
The integration over $\lambda$ in Eq.~\eqref{ccadie} is performed numerically.  $\boldsymbol{\Pi}$ is again the irreducible polarizability defined in Eq.~\eqref{pi_def}. The evaluation of Equation~\eqref{eq:sosex} scales $O(N^5)$.

In RSPT, the SE term is given by\cite{szabook}
\begin{align}
E_c^{\text{SE}} &= 
\sum_{m}^{\text{occ}}\sum_{a}^{\text{virt}} \frac{|\langle\psi_{m}|\hat{v}^{\text{HF}}-\hat{v}^{\text{MF}}|\psi_{a}\rangle|^{2}}{\epsilon_{m}-\epsilon_{a}} \label{eq:SEenergya}\\
&=\sum_{m}^{\text{occ}}\sum_{a}^{\text{virt}} \frac{|\langle\psi_{m}|\hat{f}|\psi_{a}\rangle|^{2}}{\epsilon_{m}-\epsilon_{a}} 
\label{eq:SEenergyb}
\end{align}
where $\hat{v}^{\text{HF}}$ is the HF and $\hat{v}^{\text{MF}}$ the mean-field potential from, e.g., KS-DFT. $\hat{f}$ is the Fock operator, $\hat{f}=-\nabla^{2}/2+\hat{v}^{\text{ext}} + \hat{v}^{\text{HF}}$ and $\hat{v}^{\text{ext}}$ is the external potential (electron-nuclei interaction).
The derivation of Eqs.~\eqref{eq:SEenergya} and \eqref{eq:SEenergyb} can be found in the SI of Ref.~\citenum{Ren_11_SE}. It is obvious from  Eq.~\ref{eq:SEenergya} that the SE contribution is zero if $\hat{v}^{\text{MF}}$ corresponds to the HF potential, which directly follows from Brillouin's theorem.\cite{szabook}

The renormalization of the SEs was proposed to avoid divergence problems for systems with vanishing KS gaps. The rSE approximation was initially implemented approximately,\cite{Ren_11_SE} including only the diagonal terms in the infinite expansion. The rSE approach was later refined,\cite{rPT2Ren13} accounting now also for the off-diagonal elements. The refined version is the one we use in this work. The procedure relies on a subspace diagonalization as follows: we first evaluate the occupied block $f_{mn}$ and unoccupied $f_{ab}$ of the Fock matrix with the KS single-particle orbitals $\{\psi_p\}$, i.e., $f_{mn}=\langle\psi_{m}|\hat{f}|\psi_{n}\rangle$ and $f_{ab}=\langle\psi_{a}|\hat{f}|\psi_{b}\rangle$. The two blocks $f_{mn}$ and $f_{ab}$ are then diagonalized separately which yields two new sets of eigenvalues $\tilde{\varepsilon}_{m/a}$ and orbitals $\tilde{\psi}_{m/a}$. The rSE correlation energy is then obtained by replacing the KS energies and orbitals in Eq.~\eqref{eq:SEenergyb} with the ones obtained after the subspace diagonalization yielding
\begin{align}
E_c^{\text{rSE}} 
=\sum_{m}^{\text{occ}}\sum_{a}^{\text{virt}} \frac{|\langle\tilde{\psi}_{m}|\hat{f}|\tilde{\psi}_{a}\rangle|^{2}}{\tilde{\epsilon}_{m}-\tilde{\epsilon}_{a}} 
\label{eq:rSEenergy}
\end{align}
The rSE correction was first investigated in the context of RPA,\cite{Ren_11_SE,rPT2Ren13,RPA_Ren12,rPT2_1} but has been also explored for quasiparticle energies from $GW$\cite{jinRenormalizedSinglesGreen2019,Li2022a,liRenormalizedSinglesGreen2021} and the $T-$matrix approximation,\cite{liRenormalizedSinglesGreen2021} charge-neutral excitations from BSE$@GW$\cite{Li2022b} and dissociation curves from multireference methods.\cite{liMultireferenceDensityFunctional2022}

\subsubsection{GW quasiparticle energies}

The $GW$ approximation to Hedin's equations\cite{Hedin65} is a widespread method for the computation of photoemission spectra of molecules and solids.\cite{GWcompe} It has been traditionally applied to valence excitations\cite{GWcompe} and is now also employed for deep-core levels.\cite{Golze18,Golze2020,Keller2020,Golze2022,PanadesBarrueta2023,DanielContour2,Kehry2023,ZhuContour3} Central to $GW$ is the self-energy $\Sigma$, which is obtained  from the Green's function $G$ and the screened Coulomb interaction $W$. The poles of $G$ are the quasiparticle energies (QPs), which can be directly linked to the charged excitations measured in photoemission spectroscopy. The most common flavor of $GW$ is the single-shot $G_0W_0$ approach, where the QP energies are obtained as correction to the KS-DFT eigenvalues via the QP equation:
\begin{equation}\label{qpeq}
    \epsilon_{p}^{G_0W_0} = \epsilon_{p} + \Re\left[\Sigma_p\left(\epsilon_{p}^{G_0W_0}\right)\right] - v_{p}^{xc}
\end{equation}
with $\Sigma_p = \langle\psi_p|\Sigma|\psi_p\rangle$ and $v^{xc}_p = \langle\psi_p|v^{xc}|\psi_p\rangle$ and where $v^{xc}$ denotes the exchange-correlation potential from KS-DFT. The evaluation of the  self-energy correction term $\Sigma_p(\omega)$ involves a frequency integration over $G_0$ and $W_0$, which can be performed with different techniques.\cite{GWcompe} The RI-V approach has been used in several frequency integration techniques, such as the fully-analytic approach \cite{SettenTubomole,brunevalmolgw}, contour deformation\cite{Golze18,HolzerContour1,DanielContour2,ZhuContour3} and analytic continuation.\cite{Ren12,WilhelmAC2,HolzerContour1} We use here the analytic continuation, where we calculate the self-energy for a  set of imaginary frequencies $\{\ii\omega\}$. The RI-V expression for $\Sigma_p(\ii\omega)$ reads\cite{Ren12} 
\begin{equation}\label{sigmaRI}
    \Sigma_{p}(\ii\omega) = -\frac{1}{2\pi} \sum_{q}\int_{-\infty}^{\infty} d\omega' \frac{\displaystyle \sum_{\mu\nu} O_{pq}^{\mu} [1-\Pi(\ii\omega)]_{\mu\nu}^{-1} O_{qp}^{\nu}}{\ii\omega + \ii\omega' + \epsilon_{F} -\epsilon_{m}}
\end{equation}
where $\epsilon_{F}$ is the Fermi energy and $\Pi_{\mu\nu}$ is given in Eq.~\eqref{pi_def}. The self-energy is then analytically continued from the imaginary to the real-frequency axis before the QP energies are calculated by Eq.~\eqref{qpeq}. The evaluation of Eq.~\eqref{sigmaRI} scales $O(N^4)$.

\subsection{Separable Resolution of the Identity}

In RI-V, the atomic orbital product $\varphi_{i}(\mathbf{r})\varphi_{j}(\mathbf{r})$ is entangled through the three-center Coulomb integrals in the expansion coefficients $C_{ij}^{\mu}$ (Eq.~\eqref{coeff_riv}). The idea of the separable RI or equivalently RI-RS is to disentangle this product, similarly as in ISDF, by the following separable expression \cite{BlaseRPA}
\begin{equation}\label{coeff_rirs}
    [C_{ij}^{\mu}]_{\text{RI-RS}} = \sum_{k}  \varphi_{i}(\mathbf{r}_{k}) \varphi_{j}(\mathbf{r}_{k}) Z_{k\mu}
\end{equation}
where $k$ runs over a given set of real-space grid points $\{\mathbf{r}_k\}$.
In Eq.~\eqref{coeff_rirs}, the orbital basis functions, $\varphi_{i}(\mathbf{r}_{k})$, are evaluated for each $\mathbf{r}_{k}$ grid point and the contraction coefficients $Z_{k\mu}$ are obtained by employing the minimization condition:
\begin{equation}\label{minz1}
    \argmin_{Z_{k\mu}}\sum_{ij}\left([C_{ij}^{\mu}]_{\text{RI-RS}}-[C_{ij}^{\mu}]_{\text{RI-V}}\right)^{2}.
\end{equation}
The reference for the RI-RS approach, as proposed by Duchemin and Blase,\cite{BlaseRPA} is RI-V, but could be in principle replaced by another RI flavor. In this work, we keep the RI-V reference due to its known high accuracy. 

With $D_{ij}^{k}$ defined as follows
\begin{equation}\label{bas_prod_rirs}
D_{ij}^{k}=\varphi_{i}(\mathbf{r}_{k}) \varphi_{j}(\mathbf{r}_{k})
\end{equation}
Eq. \eqref{minz1} can be written as 
\begin{equation}\label{minz2}
    \argmin_{Z_{k\mu}}\sum_{ij}\left(\sum_{k}D_{ij}^{k}Z_{k \mu}-[C_{ij}^{\mu}]_{\text{RI-V}}\right)^{2}
\end{equation}
which yields
\begin{equation}\label{z_coeff}
Z_{k \mu} = \sum_{k'}\left(\sum_{ij}D_{ij}^{k}D_{ij}^{k'}\right)^{-1} \left(\sum_{ij}D_{ij}^{k'}[C_{ij}^{\mu}]_{\text{RI-V}}\right).
\end{equation}
Following our notation for RI-V in Eq.~\eqref{fcao1}, we obtain for the  AO-based three-center quantity $M_{ij}^{\mu}$ in RI-RS:
\begin{equation}\label{fcao_rirs}
[M_{ij}^{\mu}]_{\text{RI-RS}} = \sum_{k}D_{ij}^{k} M_{k \mu} 
\end{equation}
with
\begin{equation}\label{muk_rirs}
    M_{k \mu} = \sum_{\nu}Z_{k \nu} V_{\nu\mu}^{1/2}.
\end{equation}
The MO transformation of $[M_{ij}^{\mu}]_{\text{RI-RS}}$ yields the RI-RS version of Eq.~\eqref{fcmo1}:
\begin{equation}\label{fcmo_rirs}
    [O_{p q}^{\mu}]_{\text{RI-RS}} =\sum_{ij} c_{p}^{i}c_{q}^{j} [M_{ij}^{\mu}]_{\text{RI-RS}},
\end{equation} 
which replaces the three-center quantity $O_{pq}^{\mu}$ in the RI-V expressions of the methods introduced in Section~\ref{sec:riv_methods}. For example, the Fock exchange energy in RI-RS becomes
\begin{equation}\label{Fock_ener_rirs}
    E_x^{\text{HF}} = - \sum_{mn}^{\text{occ}}\sum_{\mu}[O_{mn}^{\mu}]_{\text{RI-RS}}[O_{nm}^{\mu}]_{\text{RI-RS}}.
\end{equation}
The RI-RS expressions for MP2, CCSD, RPA, SOSEX, rSE and $GW$ are obtained analogously.  

\section{Implementation details}
\label{sec:implementation}
The separable-RI approach has been implemented in the all-electron program package FHI-aims,\cite{FHIaims} which expands the MOs in NAO basis functions  $\{\varphi_{i}\}$. The NAOs have the following form
\begin{equation}\label{naodef}
    \varphi_{i} (\mathbf{r}) = \frac{u_{i}(r)}{r}Y_{lm}(\Omega)
\end{equation}
where $u_{i}$ are radial functions and $Y_{lm}(\Omega)$ spherical harmonics. 
The radial part of the NAOs is fully flexible and not restricted to a particular shape. Popular local basis functions such as GTOs or Slater-type orbitals (STOs) present special forms of an NAO and are numerically tabulated when provided as input to FHI-aims. 

The ABFs are also NAOs in FHI-aims. Most localized-basis set codes use pre-optimized ABFs obtained via variational procedures\cite{Weigend98} whereas the default procedure in FHI-aims is to compute them on-the-fly during run-time. The ABFs are obtained by generating on-site products of the primary basis functions. Linear dependencies are removed with a Gram-Schmidt orthogonalization procedure. More details can be found in Ref.~\citenum{Ren12}.
\begin{figure}[t]
\includegraphics[]{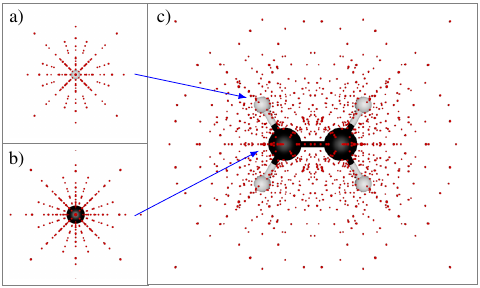}
    \caption{Real space grid ${\mathbf{r}_k}$ (red dots) for a) the hydrogen atom, b) carbon atom and c) ethylene molecule. The real space grid for a given molecule is constructed by the superposition of the grid of each species. \DG{Displayed are the grids for the cc-pVTZ basis set.}}
    \label{grids_figures}
\end{figure}

The real-space grid points $\{\mathbf{r}_k\}$ are pre-optimized for the primary basis set of each atomic species. We used in this work the $\{\mathbf{r}_k\}$ grids \DG{developed} by Duchemin and Blase,\cite{BlaseRPA,BlaseGW} \DG{which was so far restricted to the cc-pVTZ and def2-TZVP basis sets.}
The atomic grids are constructed for each atom type by different combinations of four base shells, which are subsets of Lebedev quadrature grids.\cite{LEBEDEV197544} Each base shell is replicated for each species with different radii. The number of repetitions of each base shell and the corresponding radius were optimized such that $\sum_{ij}\sum_{\mu}([C_{ij}^{\mu}]_{\text{RI-RS}}-[C_{ij}^{\mu}]_{\text{RI-V}})^{2}$ is minimized for the isolated atom case.\cite{BlaseRPA} \DG{This procedure was followed for the grids optimized for usage with the cc-pVTZ basis set.\cite{BlaseRPA} In later work by Duchemin and Blase,\cite{BlaseGW} the procedure was further refined by a subsequent optimization step, where the grid points are allowed to move freely without being constrained by any symmetry condition. The grids for the def2-TZVP basis sets were generated by this more sophisticated approach.}

Figure \ref{grids_figures} shows the $\mathbf{r}_k$ grid points \DG{for the cc-pVTZ basis set} for a) a hydrogen atom, b) a carbon atom and c) the ethylene molecule. We observe that the grids are very compact, which is essential for low-scaling algorithms with small prefactors. The number of grid points is only approximately 10 times larger than the number of primary basis functions. For the \DG{cc-pVTZ} grids used in this work, we have $\approx 310$ points for each  C, N, and O atom and 167 points for the H atom. \DG{The $\{\mathbf{r}_k\}$ grids for def2-TZVP are of similar size: 136 points for H, 100 for He and 336 for second-row elements (C, N, O etc.).\cite{BlaseGW} The number of points increases to 436 and 536 points for third and fourth-row elements, respectively.\cite{BlaseGW}} The system-specific grid points are constructed as shown in Figure~\ref{grids_figures}c: the real-space grid for the ethylene molecule is formed by a superposition of the ${\mathbf{r}_k}$ points of each hydrogen and carbon atom. \DG{The grids for the def2-TZVP basis are shown in Figure~S1 (Supporting Information), which demonstrates that the arrangement of the grids points is significantly less symmetric as the one shown for the cc-pVTZ grids in Figure~\ref{grids_figures}.}

\begin{figure}[t]
    \centering
    \includegraphics{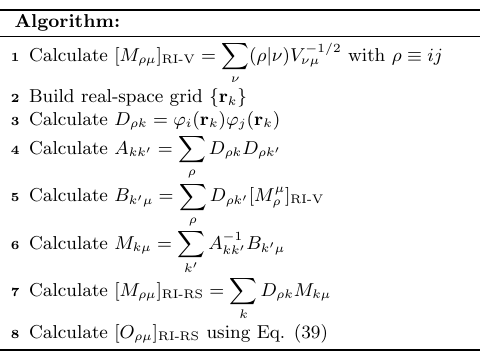}
    \caption{Algorithm for the computation of separable-RI fitting coefficients}
    \label{pseudo_rirs}
\end{figure}

The pseudocode for our RI-RS implementation in FHI-aims is summarized in Figure \ref{pseudo_rirs}. We start by computing the coefficients $M_{ij}^{\mu}$. The two-center Coulomb integrals $V_{\mu\nu}$, Eq. \eqref{2c-eris}, are computed with logarithmic Bessel transforms\cite{Talman09} in Fourier space.\cite{Talman03,Talman07} The evaluation of the three-center integrals, Eq.~\eqref{3c-eris}, is performed in real space using overlapping atom-centered spherical grids; see also Ref.~\citenum{Ren12} for more details on the NAO integration procedures. In the next step, we build the real-space grids ${\mathbf{r}_k}$ from the provided specifications of the base shells and  corresponding radii \DG{(cc-pVTZ grids) or directly read them from an external file (def2-TZVP grids).}
In step 3, the product of two basis functions, $D_{ij}^{k}$, is computed for each grid point ${\mathbf{r}_k}$. 
We only include pairs $ij$, where the atomic orbitals $\varphi_i$ and $\varphi_j$ have a significant overlap of their spatial extension.  
To simplify the computation of $D_{ij}^k$, we form the combined index $\rho\equiv ij$ yielding a matrix $\mathbf{D}$ of dimension $N_{\rho}\times N_k$, where $N_\rho$ is the number of pairs and $N_k$ the number of grid points. 
In steps 4 and 5, we build the auxiliary matrices $\mathbf{A}$ and $\mathbf{B}$, respectively and invert $\mathbf{A}$. We found that the inversion is numerically stable without $L_2$ regularization, which was proposed in Ref.~\citenum{BlaseGW}. The quantity $M_{k\mu}$, Eq. \eqref{muk_rirs}, is obtained by matrix multiplication of $\mathbf{A}^{-1}$ and $\mathbf{B}$. Finally, the RI-RS fitting coefficients (in the AO basis) are obtained from Eq.~\eqref{fcao_rirs}. Algorithm \ref{pseudo_rirs} shows that the computation of the RI-RS coefficients is straightforward, only requiring matrix multiplications and one inversion.

\section{Computational details}
\label{sec:computational_details}
\begin{figure}[t]
    \centering
    \includegraphics{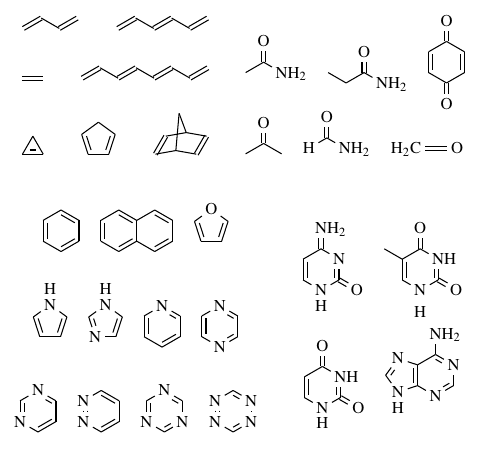}
    \caption{Thiel test set of small organic compounds}
    \label{Fig:Thiels}
\end{figure}

For the validation of the RI-RS implementation, we used the Thiel set,\cite{thielset}\DG{the GW100 benchmark set~\cite{Setten15} and 10 of the armorphous carbon clusters from our previous work.~\cite{Golze2022} The Thiel set} comprises 28 small organic molecules composed of the elements C, N, O and H.  The whole benchmark set is shown in Figure~\ref{Fig:Thiels}. We used the MP2/6-31-G(d) geometries supplied in Ref.~\citenum{thielset} by Thiel and collaborators. \DG{The GW100 benchmark set contains 100 molecules, encompassing a broad spectrum of elements from the periodic table. We used the structures reported in the original GW100 benchmark study\cite{Setten15} and exclude the seven molecules, which contain 5\textsuperscript{th} period elements (\ce{Xe}, \ce{Rb2}, \ce{I2}, \ce{C2H3I}, \ce{CI4}, \ce{AlI3}, \ce{Ag2}). For these seven molecules, optimized real-space grid points $\{\mathbf{r}_k\}$ are currently still lacking. The amorphous carbon clusters were obtained by carving clusters from periodic structures of three-dimensional disordered carbon materials with the elemental composition CHO. Dangling C-C bonds were passivated with hydrogen, see Ref.~\citenum{Golze2022} for more details.}

We used Dunning's cc-pVTZ\cite{ccpvtzbasis} basis sets \DG{for all calculations with the Thiel benchmark set and for the amorphous carbon clusters. We repeated the HF and MP2 calculations for the Thiel benchmark set with the def2-TZVP basis set.\cite{Weigend2005} We also used the def2-TZVP basis set for the GW100 benchmark. Both basis sets are spherical GTOs. For the RI-RS calculations, we used the atomic grids $\{\mathbf{r}_k\}$ specifically optimized for the respective basis set. For cc-pVTZ, we utilized the grids published in the SI of Ref.~\citenum{BlaseRPA}, while for def2-TZVP, we used the grids employed in Ref.~\citenum{BlaseGW} and published in Ref.~\citenum{zenodorepo}.}

We performed HF and MP2 calculations without RI with the NWChem code\cite{nwchemcode} to obtain exact 4c-ERIs references values. The NWChem default settings are otherwise used in these calculations. 
HF, MP2, CCSD, RPA, SOSEX, rSE, rPT2 and $GW$ are exclusively implemented with RI in the FHI-aims code. Non-RI versions of these methods are not implemented. We show in the following results from RI-V and RI-RS. For the automatic generation of the ABFs, we used a threshold of $1.0\times10^{-10}$ for the basis-basis product screening. 
The Perdew-Burke-Ernzerhof (PBE) \cite{PBE_orig,PBE_orig_erratum} exchange-correlation functional was used for the preceding KS-DFT calculations in RPA, SOSEX, rSE, rPT2 and $G_0W_0$ \DG{with the Thiel benchmark set. The GW100 calculations employed the PBE0~\cite{Adamo1999,Ernzerhof1999} hybrid functional as starting point, while the $G_0W_0$ calculations of the amorphous carbon clusters started from a PBE-based hybrid (PBEh)\cite{Atalla2013} functional with 45\% of exact exchange $\alpha$ denoted as PBEh($\alpha$=0.45).} 
The frequency integral of the $GW$ self-energy (Eq.~\eqref{sigmaRI}) was computed using a modified Gauss-Legendre grid\cite{RPA_Ren12} with 200 imaginary frequency points. The self-energy $\Sigma_n(\ii\omega)$ was computed for the same set $\{\ii\omega\}$ and then analytically continued to the real axis employing a Pad\'e model\cite{Pade_paper} with 16 parameters.  
The QP equation (Eq.~\eqref{qpeq}) was solved iteratively. The SOSEX frequency integral in Eq.~\eqref{eq:sosex} was evaluated numerically using also a modified Gauss Legendre grid with 200 points. The integration over $\lambda$ in SOSEX (Eq.~\eqref{ccadie}) was computed using a Gauss-Legendre quadrature with 5 points. We made the input and output files of all the FHI-aims calculations (RI-V and RI-RS) available in the NOMAD database.\cite{nomadrepo} 

\begin{table*}[t!]
    \centering
    \caption{Mean errors (ME), mean absolute errors (MAEs) and maximum absolute error (maxAE) with respect to the non-RI reference for total HF exchange and MP2 correlation energies [meV/electron] and HF and MP2 atomization energies [meV]. The errors are defined as $E^{\textnormal{RI-V/RI-RS}}-E^{\textnormal{non-RI}}$.}
    \label{tab:non_ri}
    \begin{tabular*}{0.99\linewidth}{@{\extracolsep{\fill}}lrrrrrrrrrrrrrrrr} \toprule
 &   \multicolumn{8}{c}{Total energy}  & \multicolumn{8}{c}{Atomization energy } \\  \cmidrule{2-9} \cmidrule{10-17}
      & \multicolumn{4}{c}{cc-pVTZ} &  \multicolumn{4}{c}{def2-TZVP} & \multicolumn{4}{c}{cc-pVTZ} &  \multicolumn{4}{c}{def2-TZVP} \\ \cmidrule{2-5} \cmidrule{6-9}  \cmidrule{10-13} \cmidrule{14-17}
            & \multicolumn{2}{c}{HF exch.} &  \multicolumn{2}{c}{MP2 corr.}  & \multicolumn{2}{c}{HF exch.} &  \multicolumn{2}{c}{MP2 corr.}  & \multicolumn{2}{c}{HF} &  \multicolumn{2}{c}{MP2} & \multicolumn{2}{c}{HF} &  \multicolumn{2}{c}{MP2}\\ \cmidrule{2-3} \cmidrule{4-5} \cmidrule{6-7} \cmidrule{8-9} \cmidrule{10-11} \cmidrule{12-13}\cmidrule{14-15}\cmidrule{16-17}   
            & RI-V   & RI-RS &  RI-V & RI-RS  & RI-V   & RI-RS &  RI-V & RI-RS &  RI-V   & RI-RS &  RI-V & RI-RS  & RI-V   & RI-RS &  RI-V & RI-RS   \\ \cmidrule{2-3} \cmidrule{4-5}\cmidrule{6-7} \cmidrule{8-9} \cmidrule{10-11} \cmidrule{12-13} \cmidrule{14-15} \cmidrule{16-17}
    ME      & -0.019 & 0.132 &  -0.003 & -0.004  & -0.022 &	0.015 & -0.002 &	-0.010  & -0.490 & -8.568 & 0.045 & -0.274 & -0.145 &	-2.997 &	0.146 &	0.531 \\  
    MAE     &  0.063 & 0.207 &   0.005 &  0.025  & 0.064	& 0.124	& 0.005	& 0.013  & 0.700 & 9.927  &  0.633 &  1.324   & 0.483	& 4.205	& 0.451	& 0.744 \\
    maxAE   &  0.175 & 0.645 &   0.016 &  0.066  & 0.178	& 0.341	& 0.018	& 0.038  & 3.962 & 29.219 &  3.010 &  8.481 & 1.706	& 11.965	& 1.857	& 3.319 \\ \bottomrule
    \end{tabular*}
\end{table*}

We computed the correlation energies ($E_c$) from MP2, CCSD, RPA, SOSEX, rSE and rPT2 and the HF exchange with RI-V or RI-RS. We also computed MP2, CCSD, RPA and rPT2 total energies ($E_t$). The MP2 and CCSD total energies were computed as 
\begin{equation}\label{etotal_mp2_ccsd}
    E_{t}^{\text{MP2/CCSD}} = E_{gs}^{\text{HF}_{\text{RI-V}}} + E_{c}^{\text{MP2/CCSD}}
\end{equation}
where $E_{gs}^{\text{HF}_{\text{RI-V}}}$ is the HF ground state energy. The RPA and rPT2 total energies were calculated as 
\begin{equation}\label{etotal_rpas}
    E_{t}^{\text{RPA/rPT2}} = E_{gs}^{\text{PBE}} - E_{xc}^{\text{PBE}}+ E_{x}^{\text{HF}_{\text{RI-V}}} + 
    E_{c}^{\text{RPA/rPT2}}
\end{equation} 
where $E_{gs}^{\text{PBE}}$ is the KS-DFT ground state with PBE, $E_{xc}^{\text{PBE}}$ is the PBE exchange-correlation energy and $E_{x}^{\text{HF}_{\text{RI-V}}}$ the HF exchange based on RI-V. Note that we computed $E_{gs}^{\text{HF}}$ and  $E_{x}^{\text{HF}}$ in Eqs.~\eqref{etotal_mp2_ccsd} and ~\eqref{etotal_rpas} always with RI-V, also when RI-RS is used for $E_c$. We used the total energies to compute atomization energies $\Delta E_{a}$ for molecules with the general formula $\text{C}_{n}\text{N}_{m}\text{O}_{l}\text{H}_{k}$ as 
\begin{equation}
\begin{split}\label{eq_atom_ener}
    \Delta E_{a}(\text{C}_{n}\text{N}_{m}\text{O}_{l}\text{H}_{k}) =& [nE_{t}(\text{C})+mE_{t}(\text{N}) + lE_{t}(\text{O}) + kE_{t}(\text{H})] \\&-E_{t}(\text{C}_{n}\text{N}_{m}\text{O}_{l}\text{H}_{k})
 \end{split}
\end{equation}
where $E_t(\text{C}_{n}\text{N}_{m}\text{O}_{l}\text{H}_{k})$ is the total energy of the molecule and $E_t(\text{C})$, $E_t(\text{N})$, $E_t(\text{O})$ and $E_t(\text{H})$ is the total energy of the carbon, nitrogen, oxygen and hydrogen atom, respectively. Since we are interested in a purely numerically comparison of RI approaches, we omitted corrections for the basis set superposition error (BSSE). We note that such BSSE corrections are necessary for a comparison to computational and experimental reference values.


\begin{figure*}[h]
\includegraphics{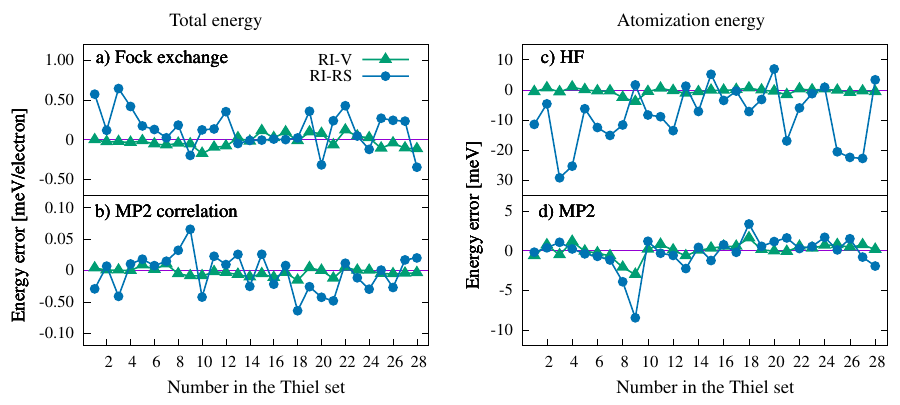}
    \caption{a) Errors of RI-V and RI-RS with respect to the non-RI reference. Total energy errors [meV/electron] for a) HF exchange and b) MP2 correlation energy. Atomization energy errors [meV] for c) HF and d) MP2. The errors are defined as $E^{\textnormal{RI-V/RI-RS}}-E^{\textnormal{non-RI}}$. \DG{The cc-pVTZ basis set was used.}}
    \label{fig:non_ri}
\end{figure*}
\section{Results and discussion}
\label{sec:results}

\subsection{Comparison to non-RI reference}
\label{subsec:results_nonRI}
We start the discussion by comparing our RI-V and RI-RS results to the exact 4c-ERIs reference (non-RI) obtained from NWChem. We restrict this comparison to HF and MP2. The errors with respect to the non-RI values are displayed in Figures~\ref{fig:non_ri} a and b for the total and in c and d for the atomization energies, respectively. The corresponding mean error (ME), mean absolute error (MAE) and maximum absolute error (maxAE) are given in Table~\ref{tab:non_ri}. \DG{The discussion refers, unless otherwise noted, to the cc-pVTZ results.}

For the total energies, we report the errors for the HF exchange $E_x^{\textnormal{HF}}$ (Eq.~\eqref{Fock_ener}) in Figure~\ref{fig:non_ri}a and the MP2 correlation energy $E_c^{\textnormal{MP2}}$ (Eq.~\eqref{mp21}) in Figure~\ref{fig:non_ri}b. The errors for the MP2 correlation energy are an order of magnitude smaller, for both RI-V and RI-RS, which is expected since the absolute values of $E_c^{\textnormal{MP2}}$ and $E_x^{\textnormal{HF}}$ differ by an order of magnitude as well. The RI-V scheme reproduces the non-RI reference almost exactly, with MAEs of 0.063 meV/electron and 0.005 meV/electron for $E_x^{\textnormal{HF}}$ and $E_c^{\textnormal{MP2}}$, respectively. Previously published RI-V results for the Thiel benchmark set are similar: Duchemin \textit{et al.}\cite{BlaseRPA} reported MAEs of 0.103~meV/electron for $E_x^{\textnormal{HF}}$ and 0.030~meV/electron  for $E_c^{\textnormal{MP2}}$ with respect to the non-RI reference. Our RI-V results are closer to the non-RI reference, which we attribute to the automatic ABF generation in FHI-aims that provides a more complete RI basis set compared to the pre-optimized ones. For cc-pVTZ, the number of auto-generated ABFs is approximately a factor of two larger than the pre-optimized ones used in Ref.~\citenum{BlaseRPA}. 

As evident from Figure~\ref{fig:non_ri}a and b, the RI-RS errors oscillate around the non-RI reference without a clear trend for an over- or underestimation. The RI-RS MAEs and maxAEs of the total energies are approximately 4 to 5 times larger than the RI-V ones. Reference~\citenum{BlaseRPA} reported only a factor of 1.5 to 2. \DG{One possible cause is the neglect of} the rotational non-invariance of the RI-RS approach. Unlike NAOs or GTOs, the $\{\mathbf{r}_k\}$ grid is not rotational invariant, which was accounted for in the previous work by averaging over 40 random orientations of each molecule in the Thiel set.\cite{BlaseRPA} However,  the effect is small. The RI-RS MAEs reported with respect to the non-RI results were 0.171 meV/electron ($E_x^{\textnormal{HF}}$) and 0.0435~mev/electron ($E_c^{\textnormal{MP2}}$) in Ref.~\citenum{BlaseRPA}. Our RI-RS MAEs are very similar or even smaller (0.207/0.025 meV/electron for HF/MP2) because our RI-V reference is closer to the non-RI result. This demonstrates that the influence of other parameters is larger, such as improving an already good RI basis set, justifying to neglect the rotational non-invariance.

Turning from total to relative energies, we compute the atomization energies according to Eq.~\eqref{eq_atom_ener}. For MP2, we compute the total exchange and correlation energy $E_t$ in Eq.~\eqref{eq_atom_ener} with Eq.~\eqref{etotal_mp2_ccsd}: the HF ground state energy is calculated with RI-V, while the MP2 correlation energy is computed either with RI-V or RI-RS. The reason for keeping RI-V for the HF contribution is that $|E_x^{\textnormal{HF}}|$ is an order of magnitude larger than $|E_c^{\textnormal{MP2}}|$ and would dominate the RI-RS error in the MP2 atomization energies, making the assessment of the RI-RS performance for correlated methods difficult.
We note that keeping RI-V in the HF exchange is also a reasonable choice in a low-scaling RI-RS implementation of a correlated method, where the purpose of the RI-RS scheme would be the scaling reduction in the calculation of the correlation energy and not in the calculation of the Fock exchange. Since we have to compute the RI-V expansion coefficients also in RI-RS, using them in $E_x^{\textnormal{HF}}$ instead of their RI-RS counterparts keeps the absolute error as small as possible.

The RI-V and RI-RS atomization energy errors with respect to the non-RI reference are shown for HF and MP2 in Figure~\ref{fig:non_ri}c and d, respectively. The RI-V approach yields again excellent results with MAEs $< 1$meV and maxAEs $<4$meV for both, HF and MP2. The RI-RS MAEs are 14 times larger for HF, but only a factor of 2 for MP2. For HF, RI-RS tends to slightly  underestimate the non-RI reference data, while the MP2 errors oscillate around the reference line following the RI-V curve. The largest RI-RS HF and RI-RS MP2 errors are 29~meV and 8~meV, respectively.  Setting the generally accepted threshold of 1~kcal/mol (= 4~kJ/mol, = 43~meV) for chemical accuracy, we find that our RI-RS errors are well within this threshold. 

\DG{We repeated the non-RI, RI-V and RI-RS calculations with the def2-TZVP basis set. The results are reported in Table~\ref{tab:non_ri} and in Figure~S2 (Supporting Information). For the RI-V results, we get almost the same errors for the total and atomization energies as with cc-pVTZ. This is not surprising since also the ABF basis set for def2-TZVP is generated automatically in FHI-aims and is thus very complete. However, the RI-RS errors with respect to the non-RI reference are noticeably smaller than for cc-pVTZ. The MAEs and maxAEs for the total HF and MP2 energies are reduced by a factor of two. Likewise, the MAEs and maxAEs for the atomization energies are 2-3 times smaller. For example, the MAE/maxAE for the HF atomization energy are 10/29~meV for cc-pVTZ, but only 4/12~meV for def2-TZVP. The reduction of the RI-RS errors might be explained with the more sophisticated procedure used to generate the def2-TZVP real-space grids. While the $\{\mathbf{r}_k\}$ grids for cc-pVTZ and def2-TZVP are of similar size, the location of the def2-TZVP grids points is, unlike for the cc-pVTZ grids, not restricted to the surface of the shells of the Lebedev grids. Our results indicate that a more flexible arrangement of the real-space grid points is key to reduce the RI-RS error without increasing the computational cost. }

\begin{figure*}[t]   \includegraphics{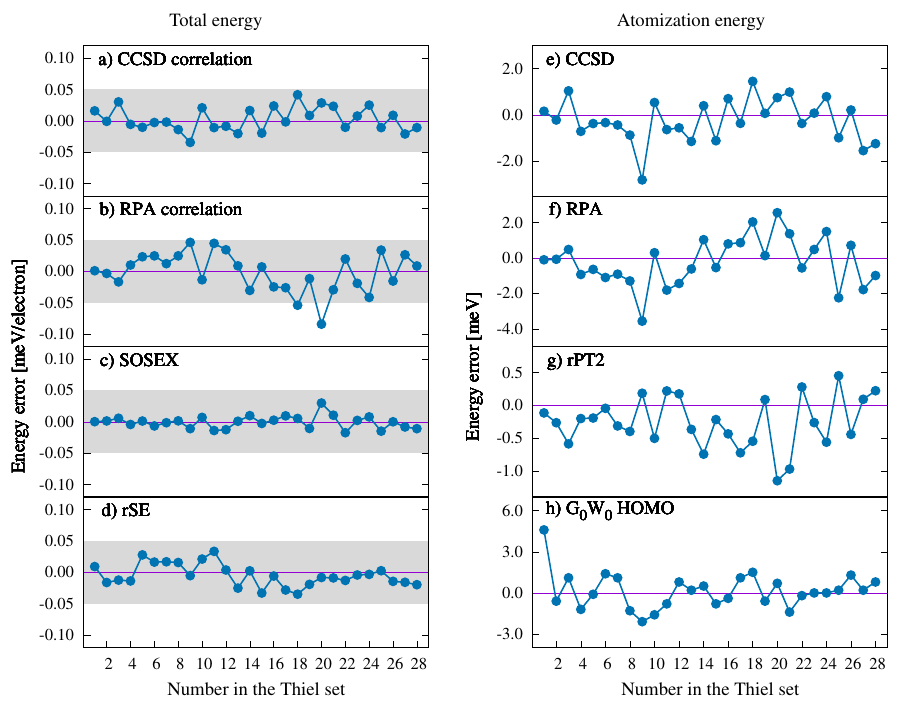}
    \caption{RI-RS errors with respect to RI-V for the total correlation energy  [meV/electron] from a) CCSD, b) RPA, c) SOSEX and d) rSE and for the atomization energy [meV] from e) CCSD, f) RPA and g) rPT2 as well as for h) the $G_0W_0$@PBE correction of the HOMO level [meV]. The errors are defined as $E^{\text{RI-RS}} - E^{\text{RI-V}}$.}
    \label{fig:ri_v}    
\end{figure*}

\subsection{Thiel benchmark for correlated methods}
\label{sec:thiel}
In this section, we extend our benchmarking of the RI-RS approach to CCSD, RPA, SOSEX, rSE, rPT2 and $G_0W_0$. We compare from now on to RI-V, which  reproduces the non-RI results almost exactly, in particular in combination with our large auto-generated auxiliary basis sets as shown in Section~\ref{subsec:results_nonRI}. To the best of our knowledge, RI-RS benchmarks for CCSD, SOSEX, rSE and rPT2 have not been presented previously, while  RI-RS results have been published for low-scaling RPA\cite{BlaseRPA} and low-scaling $G_0W_0$\cite{BlaseGW} implementations. Note that we discuss here the accuracy of RI-RS in conventional RPA and $G_0W_0$ with $O(N^4)$ scaling, excluding any other sources for possible deviations caused by a low-scaling reformulation. The errors reported in the following are exclusively due to changing the RI scheme from RI-V to RI-RS. 

In Figure~\ref{fig:ri_v}a-d we report total (correlation) energies from CCSD, RPA, SOSEX and rSE. Overall, the trend and magnitude of the errors are similar to the MP2 correlation energies discussed in the previous section. The errors are all within $\pm 0.050$~meV/electron. Only one molecule (\#20, acetone), shows a larger error of 0.084 meV/electron in the RPA correlation (see Figure~ \ref{fig:ri_v}b and Table~\ref{tab:ri_v}). The same molecule causes also the largest error (0.030 meV/electron) in the SOSEX energy. 

\begin{table}[hb]
    \centering
\caption{RI-RS mean errors (ME), mean absolute errors (MAEs) and maximum absolute error (maxAE) with respect to RI-V for total correlation energies from CCSD, RPA, SOSEX, rSE [meV/electron] and for atomization energies from CCSD, RPA, rPT2 [meV] as well as for the HOMO energies from $G_0W_0$ [meV]. The errors are defined as $E^{\text{RI-RS}}  - E^{\text{RI-V}}$.
    }
    \label{tab:ri_v}
     \begin{tabular*}{0.99\linewidth}{@{\extracolsep{\fill}}lrrrrr} \toprule
            & \multicolumn{5}{c}{Total correlation energy}   \\ \cmidrule{2-6}
           &MP2 & \ CCSD  & RPA \  & SOSEX &  rSE \\ \cmidrule{2-6}
    ME     & -0.001 & -0.002 & -0.002 & -0.001 &  -0.005 \\  
    MAE     & 0.024 & 0.015 & 0.025 & 0.007 &  0.015  \\
    maxAE   & 0.074 & 0.042 &  0.084 & 0.030 &  0.035  \\ \cmidrule{1-6}
    & \multicolumn{4}{c}{Atomization energy}  &  $G_0W_0$\ \  \\ \cmidrule{2-5} 
           &MP2 & CCSD & RPA & rPT2 & HOMO \\ \cmidrule{2-5} \cmidrule{6-6}        
    ME      & -0.319 & -0.228 & -0.225 & -0.261 &  0.157 \\  
    MAE     &  1.113 &  0.744 &  1.103 &  0.385 &  0.950 \\
    maxAE   & 5.471 &  2.799 &  3.563 &  1.151 &  4.600 \\\bottomrule  
    \end{tabular*}
\end{table}

We then turn to the RI-RS errors of the atomization energies from CCSD, RPA and rPT2 displayed in Figure~\ref{fig:ri_v}e-g. Following the same reasoning as for the MP2 atomization energies, we compute the Fock exchange in the total exchange and correlation energies $E_t^{\text{RPA,rPT2}}$ (Eq.~\eqref{etotal_rpas}) always with RI-V and only the correlation parts with RI-V or RI-RS. The errors are within 4~meV for CCSD and RPA and even within 1.2~meV for rPT2, which is an order of magnitude smaller than the threshold required for chemical accuracy. The MAEs are smaller or around 1~meV, which confirms the excellent accuracy of the RI-RS approach over a wide range of correlated methods.

Finally, we present $G_0W_0$ QP energies for the highest occupied molecular orbital (HOMO) in Figure~\ref{fig:ri_v}h. The errors are within 5~meV and the MAE is $<1$~meV, see also Tab.~\ref{tab:ri_v}. The error introduced by RI-RS is thus negligible for $GW$, where our target accuracy is usually 50~meV for comparisons between different codes. Better agreements can be reached  using the same basis set, even if the $GW$ implementations differ significantly. For example, an MAE of 3~meV has been reported comparing GW100 benchmark results of the HOMO from FHI-aims and Turbomole with the def2-QZVP Gaussian basis set.\cite{Setten15} The RI-RS error is still three times smaller.

\begin{table}[b]
    \centering
    \caption{RI-RS mean errors (ME), mean absolute errors (MAEs) and maximum absolute error (maxAE) with respect to RI-V for QP energies from $G_0W_0$@PBE0 for the GW100 benchmark set and from $G_0W_0$@PBEh($\alpha=0.45$) for amorphous carbon clusters [meV]. The errors are defined as $\epsilon^{\text{RI-RS}}  - \epsilon^{\text{RI-V}}$.}
    \label{tab:gw100_ac}
    \begin{tabular*}{0.99\linewidth}{@{\extracolsep{\fill}}lrrrr} \toprule
    \multicolumn{5}{c}{QP energy} \\ \midrule
            & \multicolumn{2}{c}{GW100} &  \multicolumn{2}{c}{Carbon clusters} \\ \cmidrule{2-3} \cmidrule{4-5}
            & HOMO   & LUMO &  HOMO & LUMO   \\ \cmidrule{2-3} \cmidrule{4-5} 
    ME      & 0.222 & -0.006  & -0.140  & -0.110 \\  
    MAE     & 0.562 & 0.212 & 0.480   &  0.180  \\
    maxAE   & 4.500 & 4.700 &  1.300  &  0.600  \\ \bottomrule
    \end{tabular*}
\end{table}

\begin{figure}[b]
    \centering
    \includegraphics[width=0.48\textwidth]{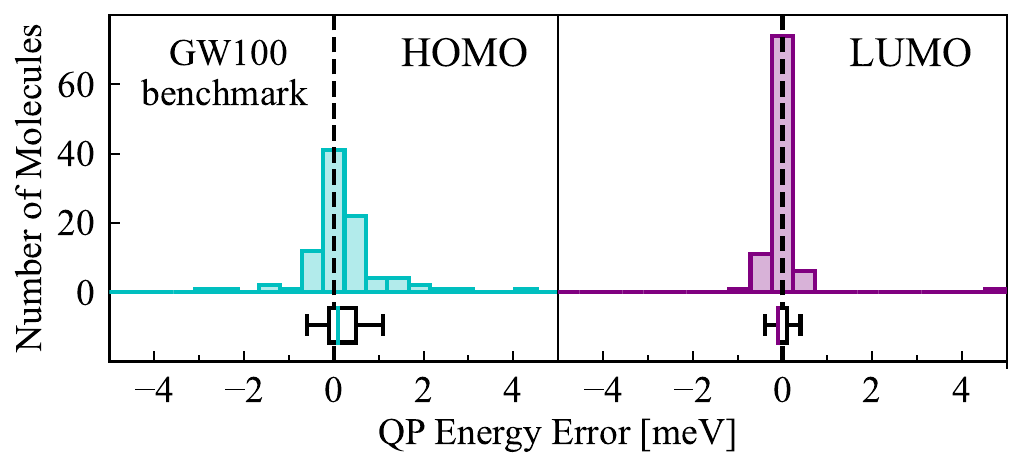}
    \caption{Histograms of QP energy errors of RI-RS with respect to RI-V at the $G_0W_0@$PBE0 level using the def2-TZVP basis. The error is defined as $\epsilon^{\text{RI-RS}} - \epsilon^{\text{RI-V}}$. The corresponding box plots are shown below the histograms.}
    \label{fig:gw100}
\end{figure}

\begin{figure}[t]
    \centering
    \includegraphics[width=0.48\textwidth]{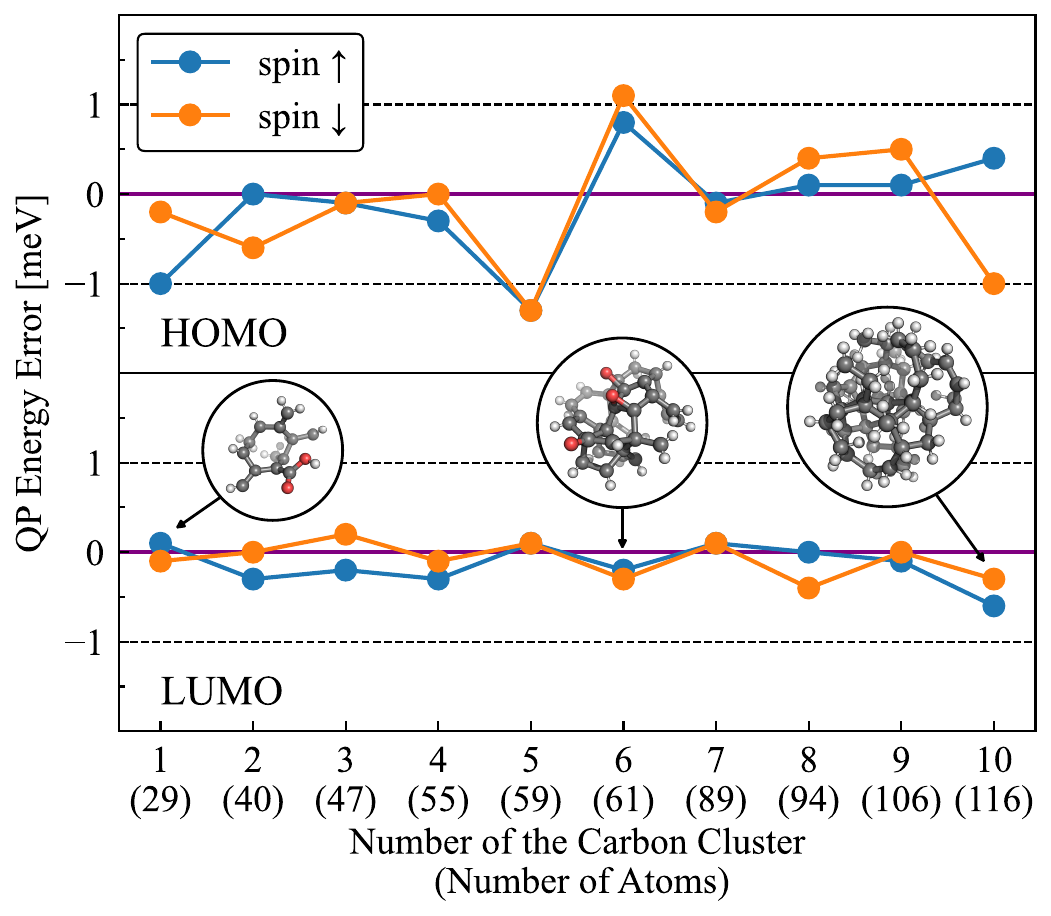}
    \caption{RI-RS error for the QP energies of amorphous carbon clusters. The QP error is defined as $\epsilon^{\text{RI-RS}} - \epsilon^{\text{RI-V}}$. The results were obtained at the $G_0W_0$@PBEh($\alpha=0.45$) level. }
    \label{fig:carbon_cluster}
\end{figure}

\subsection{QP energies for GW100 test set and carbon clusters}

\DG{For a more comprehensive assessment of the RI-RS approach, we applied it to the  GW100 benchmark set. While the Thiel set comprises only organic molecules, the GW100 test set also contains inorganic molecules, noble gases and generally also heavier elements, covering larger regions of the chemical space. As in Section~\ref{sec:thiel}, we compare the RI-RS results to the RI-V reference and use RI-RS in combination with a conventional $GW$ implementation. The QP energy errors for the HOMO and the lowest unoccupied molecular orbital (LUMO) are shown in Figure~\ref{fig:gw100} and Table~\ref{tab:gw100_ac}. As for the Thiel set, we obtain excellent results: The RI-RS errors are tightly centered around zero. The MAE is in the sub-meV range and the maxAE is below 5~meV. The errors for the LUMO are smaller, which must be attributed to the fact that their absolute values are for most systems smaller than for the HOMO.}

\DG{The GW100 benchmark was previously~\cite{BlaseGW} performed with a low-scaling $GW$ implementation based on RI-RS and the space-time method, using the same primary basis set, real-space grids and starting point. The MAEs reported in Ref.~\citenum{BlaseGW} (0.21 meV for HOMO and 0.09~meV for LUMO) are 2-3 smaller than ours, see Table~\ref{tab:gw100_ac}. It might seem surprising that the low-scaling RI-RS $GW$ implementation,\cite{BlaseGW} which is also affected by other error sources such as the time-frequency transformations, yields smaller errors than our RI-RS implementation with conventional scaling. The neglect of the rotational invariance can be excluded as possible reason since it was also not accounted for in Ref.~\citenum{BlaseGW}. The most likely cause are the different RI-V references: Ref.~\citenum{BlaseGW} uses a pre-optimized ABFs for RI-V, while our ABFs are generated automatically. A more detailed analysis shows that our auto-generated ABF set is larger and contains more diffuse functions with higher maximal angular momentum number. Therefore, our RI-V reference is closer to the RI-free results, as already discussed in Section~\ref{subsec:results_nonRI}. However, it might pose a greater challenge to reproduce our RI-V reference with RI-RS, leading to slightly larger RI-RS errors.} 

\DG{Comparing our QP-HOMO results for the Thiel and GW100 benchmark, we find that the MAE is 1.7$\times$ smaller for the latter (0.95 vs 0.56~meV). This must be attributed to the quality of the $\{\mathbf{r}_k\}$ grids: For the Thiel benchmark we used the cc-pVTZ basis set and for the GW100 the def2-TZVP basis set, where the corresponding grids were generated with more refined procedure than for cc-pVTZ. In combination with the findings in Section~\ref{subsec:results_nonRI}, the GW100 results support the assumption that lifting symmetry constraints for the real-space grids improves the accuracy of the RI-RS approach.}

\DG{So far the discussion has been restricted to small molecules. We now expand our benchmarking efforts to QP energies of large, three-dimensional amorphous carbon clusters, motivated by our primary goal to use the RI-RS approach for low-scaling $GW$ algorithms. We selected 10 clusters, which were part of our $GW$ training set used to develop machine-learning models for the prediction of X-ray photoemission spectra of disordered carbon materials. Unlike in our previous work,\cite{Golze2022} where we computed the carbon 1s excitation of the central carbon atom in each cluster, we present here the HOMO and LUMO energies. We selected amorphous carbon clusters of different sizes, ranging from 29 atoms to 116 atoms, which corresponds to 646 and 2360 basis functions, respectively. Three of the cluster are exemplarily depicted in Figure~\ref{fig:carbon_cluster}. Five clusters have the elemental composition CHO, while the other five contain only carbon and hydrogen. }

\DG{The RI-RS error of the QP energies are shown in Figure~\ref{fig:carbon_cluster}. The amorphous carbon structures are spin-polarized and the error is thus depicted for each spin channel separately. For the HOMO, we find that the error is $\leq 1$~meV for 8 of the 10 structures and the maxAE is only 1.3~meV. The error for the LUMO is even smaller with a maxAE of only 0.6~meV. The MAE and ME are averaged over both spin channels and shown in Tab.~\ref{tab:gw100_ac}. Both errors are smaller or in the same range compared to MAEs and MEs for the Thiel benchmark and the GW100 set.}

\subsection{Performance}
Replacing RI-V in a conventional implementation by RI-RS as shown in Eq.~\eqref{Fock_ener_rirs}, actually increases the computational cost. The purpose of this paper is to assess the accuracy of the RI-RS approach, but the goal is to utilize the RI-RS idea to formulate accurate low-scaling algorithms for correlated methods. For the development of low scaling algorithms, it is essential to monitor the overhead of the algorithm, i.e., the prefactor introduced by the additional operations in the computation. A large prefactor in the low-scaling algorithm might otherwise shift the crossover point, where the low-scaling implementation becomes computationally more efficient than the conventional implementation, to large system sizes, rendering the low-scaling approach in the worst case useless. The computation of the RI-RS expansion coefficients $[C^{\mu}_{ij}]_{\text{RI-RS}}$ is a step, which introduces such an additional prefactor compared to the conventional implementation: in addition to the computation of the RI-V expansion coefficients $[C^{\mu}_{ij}]_{\text{RI-V}}$, a least-square fitting is performed, which involves matrix-matrix multiplications and a matrix inversion. 

\begin{figure}[t]
    \centering
    \includegraphics{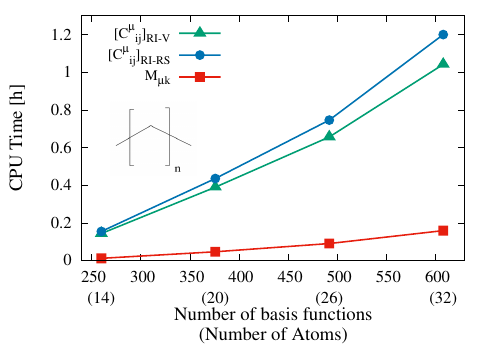}
    \caption{CPU timings [h] for computing the RI-V and RI-RS expansion coefficients for linear alkane chains. The computation of the RI-RS expansion coefficients includes the time for constructing the RI-V expansion coefficients and the computation of $M_{\mu k}$ given in Eq.~\eqref{muk_rirs}. To guide the eye, the data points are connected by lines.}
    \label{fig:timings}
\end{figure}

Figure~\ref{fig:timings} shows the CPU time for computing the RI-RS expansion coefficients for linear alkane chains C$_n$H$_{2n+2}$ with $n = 4-10$. The RI-RS time includes the computation of the RI-V expansion coefficients, which dominates the timings. Only 6 to 8 \% of the total time is spent in  calculating the coefficients $M_{\mu k}$ given in Eq.~\eqref{muk_rirs} (fitting step). The total time to compute  $[C^{\mu}_{ij}]_{\text{RI-RS}}$ also includes step 7 in the Figure \ref{pseudo_rirs}. However, for cubic-scaling RPA and $GW$ methods, the computation of $[C_{ij}^{\mu}]_{\text{RI-RS}}$ is not required, we only need $M_{\mu k}$. \cite{BlaseRPA,BlaseGW} Compared to a canonical implementation, we expect therefore a small overhead of less than 10\% in the RI step.

\section{Conclusions}
\label{sec:conclusion}

We presented an implementation of the RI-RS approach in an NAO framework. We benchmarked the accuracy of the RI-RS scheme for HF, MP2, CCSD, RPA, SOSEX, rSE, rPT2 and $G_0W_0$ by replacing the RI-V expansion coefficients with the RI-RS ones in conventional algorithms of the methods. While this strategy does not result in a speed-up of the calculation, it allows a direct assessment of the RI-RS errors. 

We computed total and atomization energies for the 28 molecules of the Thiel benchmark set. \DG{We presented QP energies at the $G_0W_0$ level for the Thiel and GW100 benchmark sets as well as large amorphous carbon clusters up to 116 atoms.} We showed that RI-V in combination with our auto-generated ABFs reproduces the non-RI calculation de-facto exactly and thus set RI-V as reference for our RI-RS benchmarks. We found that RI-RS reproduces the RI-V results for atomization energies from MP2, CCSD, RPA and rPT2 as well as $G_0W_0$ QP energies on average within 1~meV or better. The RI-RS error of the HF atomization energies is as expected one order of magnitude larger. \DG{We also showed that the RI-RS error remains in the sub-meV range, independent of the system size.}

We also found that the RI-RS scheme only introduces a small additional computational prefactor of less than 10\% in the RI step. This highlights the potential of the RI-RS approach for deriving highly accurate and computationally efficient low-scaling algorithms of correlated electronic-structure methods. Future and on-going work comprises the extension of RI-RS to periodic systems as well as the development of the actual low-scaling algorithms. Modifications of the scheme, such as fitting the RI-RS expansion coefficients to other RI flavors than RI-V, are also explored to increase the computational efficiency. A tool to generate the real-space grid points $\{\mathbf{r}_k\}$ on the fly or with little effort prior to the calculations is also necessary to exploit the potential of the RI-RS scheme.

\section*{Acknowledgement}
The authors acknowledge the European Union’s Horizon 2020 research and innovation program for financial support under the grant number 951786 (Nomad Center of Excellence) and the Horizon Europe MSCA Doctoral network grant n.101073486 (EUSpecLab). D. G. acknowledges funding by the Emmy Noether Program of the German Research Foundation (project number 453275048). We wish to acknowledge CSC – IT Center for Science (Finland), \DG{the Jülich Supercomputer Computer Center (Germany)} and Aalto Science-IT project for computational resources.

\section*{Supplementary material}
We include the computed total energies (Tables S1-S5, S10-S12), atomization energies (Tables S6, S7, and S13) and $G_0W_0$ energies (Table S8) for the Thiel set, the GW100 set (Table S14) and amorphous carbon clusters (Table S9) as well as further details on the RI-RS workflow (Section S5). \DG{A plot of the def2-TZVP real-space grids is shown for ethylene in Figure S1. RI-RS errors for HF and MP2 total and atomization energies employing the def2-TZVP basis set are plotted in Figure S2.}

\section*{Data Availability}

The data supporting the findings of this study are openly available in the NOMAD database, see Ref.~\citenum{nomadrepo}. 

\section*{References}
\bibliography{biblio}

\end{document}